\documentclass[aps]{revtex4}
\usepackage{amsmath}
\usepackage{amsfonts}
\usepackage{amssymb,epsf}
\usepackage{graphicx}

\begin{document}

\title{Thermodynamics and thermal stability of BTZ-ModMax black holes}
\author{Behzad Eslam Panah$^{1,2,3}$\footnote{
email address: eslampanah@umz.ac.ir} }
\affiliation{$^{1}$ Department of Theoretical Physics, Faculty of Science, University of
Mazandaran, P. O. Box 47416-95447, Babolsar, Iran}
\affiliation{$^{2}$ ICRANet-Mazandaran, University of Mazandaran, P. O. Box 47416-95447
Babolsar, Iran}
\affiliation{$^{3}$ ICRANet, Piazza della Repubblica 10, I-65122 Pescara, Italy}

\begin{abstract}
Motivated by a new interesting nonlinear electrodynamics (NLED) model which
is known as Modification Maxwell (ModMax) theory, we obtain an exact
analytic BTZ black hole solution in the presence of a new NLED model and the
cosmological constant. Then, by considering the obtained solution, we obtain
Hawking temperature, entropy, electric charge, mass, and electric potential.
We extract the first law of thermodynamics for the BTZ-ModMax black hole. We study thermal stability by evaluating the heat capacity (local
stability) and Helmholtz free energy (global stability). By comparing the
local and global stabilities, we find the common areas that satisfy the
local and global stabilities, simultaneously.
\end{abstract}

\maketitle

\section{Introduction}

The nonlinear electrodynamics (NLED) theories are generalizations of
Maxwell's theory and describe some of the phenomena that Maxwell's theory is
unable to explain carefully. For example, NLED theories can provide a
explanation of the self-interaction of virtual $e-e^{+}$ (i.e.
electron-positron) pairs \cite{EH,Schwinger1951,Yajima2001}. Moreover, NLED
theories can affect the gravitational redshift around super-strong
magnetized compact objects \cite{Ibrahim2002,Mosquera2004}. Applying NLED in
cosmology, we can eliminate the Big Bang's singularity. For black hole, the
singularity of spacetime remove by using NLED (see Refs. \cite%
{BBSingI,BBSingII,BBSingIII,BBSingIV}, for more details). In this regard, a
few NLED theories were introduced such as Born-Infeld (BI) \cite{BI},
Euler-Heisenberg (EH) \cite{EH}, and Power-Law (PL) \cite{PMI,PMII}. BI-NLED
preserves the electromagnetic duality invariance of Maxwell's theory and
solves the self-energy of point particles's divergence \cite{BI}. However,
BI-NLED theory is not conformally invariant. EH-NLED theory explains the
effect of vacuum polarization, but it is not conformal invariant and dual
invariant \cite{EH}. PL-NLED is conformal invariance \cite{PMI,PMII} and
also can removes the self-energy of point particle's divergence \cite%
{PMIII,PMIV}. Recently, Bandos et al. proposed an NLED theory with two
fundamental symmetries, i.e., electromagnetic duality and conformal
invariance \cite{ModMaxI}. This NLED theory had two types of solutions. One
of them leads to Bialynicki-Birula electrodynamics and another one yields a
generalization of Maxwell electrodynamics, which is known as modification of
Maxwell (ModMax) theory \cite{ModMaxI}. Notably, the ModMax NLED theory is
characterized by a dimensionless parameter $\gamma $, and it turns to
Maxwell's theory for $\gamma =0$ \cite{ModMaxI,ModMaxII} (see Refs. \cite%
{ModMaxIV,ModMaxV,ModMaxVI} for recent extensions of the ModMax NLED\
theory). By coupling the ModMax electrodynamics and gravity,
different black hole solutions have been studied in some literature \cite%
{MMBH1,MMBH2,MMBH3,MMBH4,MMBH5,MMBH6,MMBH7,MMBH9,MMBH9b,MMBH9c,MMBH10,MMBH14,MMBH15,MMBH16,MMBH17}.

The first black hole solution was found by Banados, Teitelboim, and Zanelli
in three-dimensional spacetime in Ref. \cite{BTZ}, which is famous as BTZ
black hole. Then, the study of gravity in three-dimensional spacetime
absorbed a lot of attentions due to different aspects of their physics, for
example; there are relations between effective action in string theory and
three-dimensional black holes \cite{BTZ1,BTZ2,BTZ3}, anti-Hawking phenomena 
\cite{BTZI1,BTZI2}, provides a better understanding of gravitational systems
in three-dimensional spacetime. \cite{Witten}, AdS/CFT correspondence \cite%
{Emparan2000,Carlip2005}, quantum aspect, entanglement, and quantum entropy 
\cite{Three1,Three2,Three3,Three4}, and holographic aspects \cite%
{three1,three2,three3}. In this regard, different three-dimensional black
holes have been obtained in Einstein's gravity and also modified theories of
gravity which are coupled with linear and nonlinear matters \cite%
{BTZBH1,BTZBH2,BTZBH3,BTZBH5,BTZBH6,BTZBH7,BTZBH7b,BTZBH8,BTZBH11,BTZBH12,BTZBH13,BTZBH15,BTZBH16,BTZBH17,BTZBH18,BTZBH19}.

In this paper, an analytic BTZ black hole solution is obtained by coupling Einstein's gravity with the ModMax NLED field. Then, the Hawking temperature, entropy, electric charge, electric potential, and mass of the BTZ-ModMax black hole are calculated. Finally, for evaluating the thermal stability of this black hole, the heat capacity and Helmholtz free energy are studied.

\section{Field equation and black hole solution}

The action of Einstein's theory of gravity coupled with the ModMax NLED and
the cosmological constant in three-dimensional spacetime is 
\begin{equation}
\mathcal{I}=\frac{1}{16\pi }\int_{\partial \mathcal{M}}d^{3}x\sqrt{-g}\left[
R-2\Lambda -4\mathcal{L}\right] ,  \label{Action}
\end{equation}%
where $R$ is the Ricci scalar, and $\Lambda $ is the cosmological constant.
In the above action, $g=det(g_{\mu \nu })$ devotes to the determinant of
metric tensor ($g_{\mu \nu }$). In addition, $\mathcal{L}$ is the ModMax
Lagrangian \cite{ModMaxI,ModMaxII}. Here, we assume that the ModMax Lagrangian in three-dimensional spacetime resembles the ModMax Lagrangian in four-dimensional spacetime. i.e., 
\begin{equation}
\mathcal{L}= \mathcal{S}\cosh \gamma -\sqrt{\mathcal{S}^{2}+%
\mathcal{P}^{2}}\sinh \gamma ,  \label{ModMaxL}
\end{equation}%
where $\gamma $ is a dimensionless parameter known as the ModMax parameter..
In the ModMax Lagrangian, $\mathcal{S}$ and $\mathcal{P}$, respectively, are
a true scalar, and a pseudoscalar. They are defined 
\begin{eqnarray}
\mathcal{S} &=&\frac{\mathcal{F}}{4}, \\
&&  \notag \\
\mathcal{P} &=&\frac{\widetilde{\mathcal{F}}}{4},
\end{eqnarray}%
where $\mathcal{F}=F_{\mu \nu }F^{\mu \nu }$ is the Maxwell invariant. Also, 
$F_{\mu \nu }$ is called the electromagnetic tensor field and is given as 
\begin{equation*}
F_{\mu \nu }=\partial _{\mu }A_{\nu }-\partial _{\nu }A_{\mu },
\end{equation*}%
in which $A_{\mu }$ is the gauge potential. In addition, $\widetilde{%
\mathcal{F}}=$ $F_{\mu \nu }\widetilde{F}^{\mu \nu }$, and $\widetilde{F}%
^{\mu \nu }=\frac{1}{2}\epsilon _{\mu \nu }^{~~~\rho \lambda }F_{\rho
\lambda }$. It is notable that, the ModMax Lagrangian turns to linear
Maxwell's theory, i.e., $\mathcal{L}=\frac{\mathcal{F}}{4}$, when $\gamma =0$%
.

In this paper, we want to extract the electrically charged BTZ-ModMax black
hole solutions in Einstein-$\Lambda $ gravity, so $\mathcal{P}=0$. The
Einstein-$\Lambda $-ModMax equations are written in the following form 
\begin{eqnarray}
G_{\mu \nu }+\Lambda g_{\mu \nu } &=&8\pi \mathrm{T}_{\mu \nu },  \label{eq1}
\\
&&  \notag \\
\partial _{\mu }\left( \sqrt{-g}\widetilde{E}^{\mu \nu }\right)  &=&0,
\label{eq2}
\end{eqnarray}%
where 
\begin{eqnarray}
8\pi \mathrm{T}^{\mu \nu } &=&2\left( F^{\mu \sigma }F_{~~\sigma }^{\nu
}e^{-\gamma }\right) -2e^{-\gamma }\mathcal{S}g^{\mu \nu },  \label{eq3} \\
&&  \notag \\
\widetilde{E}_{\mu \nu } &=&\frac{\partial \mathcal{L}}{\partial F^{\mu \nu }%
}=2\left( \mathcal{L}_{\mathcal{S}}F_{\mu \nu }\right) ,  \label{eq4}
\end{eqnarray}%
and $\mathcal{L}_{\mathcal{S}}=\frac{\partial \mathcal{L}}{\partial \mathcal{%
S}}$. For charged case, the ModMax field equation (Eq. (\ref{eq2})) turns to 
\begin{equation}
\partial _{\mu }\left( \sqrt{-g}e^{-\gamma }F^{\mu \nu }\right) =0.
\label{Maxwell Eq}
\end{equation}

Notably, according to the physical restriction, we consider $\gamma \geq 0$ 
\cite{ModMaxI}.

Here, we consider a three-dimensional static spacetime as 
\begin{equation}
ds^{2}=-\psi \left( r\right) dt^{2}+\frac{dr^{2}}{\psi (r)}+r^{2}d\varphi
^{2},  \label{metric}
\end{equation}%
and $\psi (r)$ is a function (known as the metric function) that we want to
find.

To extract a radial electric field, we suppose $A_{\mu }$ in the following
form 
\begin{equation}
A_{\mu }=h(r)\delta _{\mu }^{t},  \label{gauge
potential}
\end{equation}%
where by considering the metric (\ref{metric}) and the MaxMax field equation
(\ref{Maxwell Eq}), we can get 
\begin{equation}
h^{\prime }(r)+rh^{\prime \prime }(r)=0,  \label{heq}
\end{equation}%
and the prime is related to the first derivative with respect to $r$, and
the double prime is devoted to the second derivative with respect to $r$.
Using Eq. (\ref{heq}), we extract 
\begin{equation}
h(r)=q\ln \left( \frac{r}{r_{0}}\right) ,  \label{h(r)}
\end{equation}%
in which $q$ is an integration constant. This constant ($q$) is related to
the electric charge. Also, $r_{0}$ is another constant with length dimension
and confirms that the logarithmic argument is dimensionless. The obtained
electric field of ModMax NLED in three-dimensional spacetime is 
\begin{equation}
E(r)=\frac{q}{r}e^{-\gamma }.
\end{equation}

To get the metric function, i.e., $\psi (r)$, we use Eq. (\ref{metric}) with
Eqs. (\ref{eq1}), (\ref{eq3}) and (\ref{h(r)}). We can extract the following
differential equations 
\begin{eqnarray}
&&eq_{tt}=eq_{rr}=\psi ^{\prime }(r)+2\Lambda r+\frac{2q^{2}}{r}e^{-\gamma
}=0,  \label{eqENMMax1} \\
&&  \notag \\
&&eq_{\varphi \varphi }=\psi ^{\prime \prime }(r)+2\Lambda -\frac{2q^{2}}{%
r^{2}}e^{-\gamma }=0,  \label{eqENMMax2}
\end{eqnarray}%
which $eq_{tt}$, $eq_{rr}$, and $eq_{\varphi \varphi }$, respectively, are
belonged to $tt$, $rr$, and also $\varphi \varphi $ components of field
equation of motion (Eq. (\ref{eq1})). Using Eqs. (\ref{eqENMMax1}) and (\ref%
{eqENMMax2}), we obtain the metric function as 
\begin{equation}
\psi (r)=-m_{0}-\Lambda r^{2}-2q^{2}e^{-\gamma }\ln \left( \frac{r}{r_{0}}%
\right) ,  \label{f(r)EMMax}
\end{equation}%
where $m_{0}$ is an integration constant. This constant is related to the
total mass of black holes. Moreover, the obtained metric function (\ref%
{f(r)EMMax}) turns to BTZ black holes in Einstein-$\Lambda $-Maxwell theory
when the parameter of ModMax is zero, i.e. $\gamma =0$.

After finding the solution (\ref{f(r)EMMax}), we want to evaluate the
existence of essential singularity(ies). For this purpose, we evaluate two
famous scalars which are the Ricci and Kretschmann scalars. These scalars,
respectively, are 
\begin{eqnarray}
R &=&6\Lambda +\frac{2q^{2}}{r^{2}}e^{-\gamma }, \\
&&  \notag \\
R_{\alpha \beta \gamma \delta }R^{\alpha \beta \gamma \delta } &=&12\Lambda
^{2}+\frac{8\Lambda q^{2}}{r^{2}}e^{-\gamma }+\frac{12q^{4}}{r^{4}}%
e^{-2\gamma }.
\end{eqnarray}%
Our results indicates that there are different behavior for the Ricci and
Kretschmann scalars. These are;

i) for the finite and small value of $\gamma $, these quantities reveal that
there is a curvature singularity at $r=0$.

ii) in the limit of $r\longrightarrow \infty $, the Ricci and Kretschmann
scalars turn to $6\Lambda $ and $12\Lambda $, respectively, which indicate
that the asymptotical behavior of the solution is (A)dS for $\Lambda >0$ ($%
\Lambda <0$).

iii) for very large value of $\gamma $ or in the limit $\gamma
\longrightarrow \infty $, by considering $r\longrightarrow 0$, we have 
\begin{eqnarray}
\underset{r\longrightarrow 0}{\lim }R &\rightarrow &6\Lambda , \\
&&  \notag \\
\underset{r\longrightarrow 0}{\lim }R_{\alpha \beta \gamma \delta }R^{\alpha
\beta \gamma \delta } &\rightarrow &12\Lambda ^{2},
\end{eqnarray}%
where indicates that they cannot reveal a curvature singularity at $r=0$.

To find a singularity of the obtained solution (\ref{f(r)EMMax}), we
calculate some of the components of Ricci and Riemannian tensors, which are 
\begin{eqnarray}
R^{\varphi \varphi } &=&\frac{2\Lambda r^{2}+2q^{2}e^{-\gamma }}{r^{4}}, \\
&&  \notag \\
R^{r\varphi r\varphi } &=&\frac{\psi (r)\left( \Lambda r^{2}+q^{2}e^{-\gamma
}\right) }{r^{4}},
\end{eqnarray}%
and diverge in the limit $\gamma \longrightarrow \infty $, and $r\rightarrow
0$. Therefore, an essential curvature singularity exists at $r=0$, for
different values of $\gamma $.

\begin{figure}[tbph]
\centering
\includegraphics[width=0.4\linewidth]{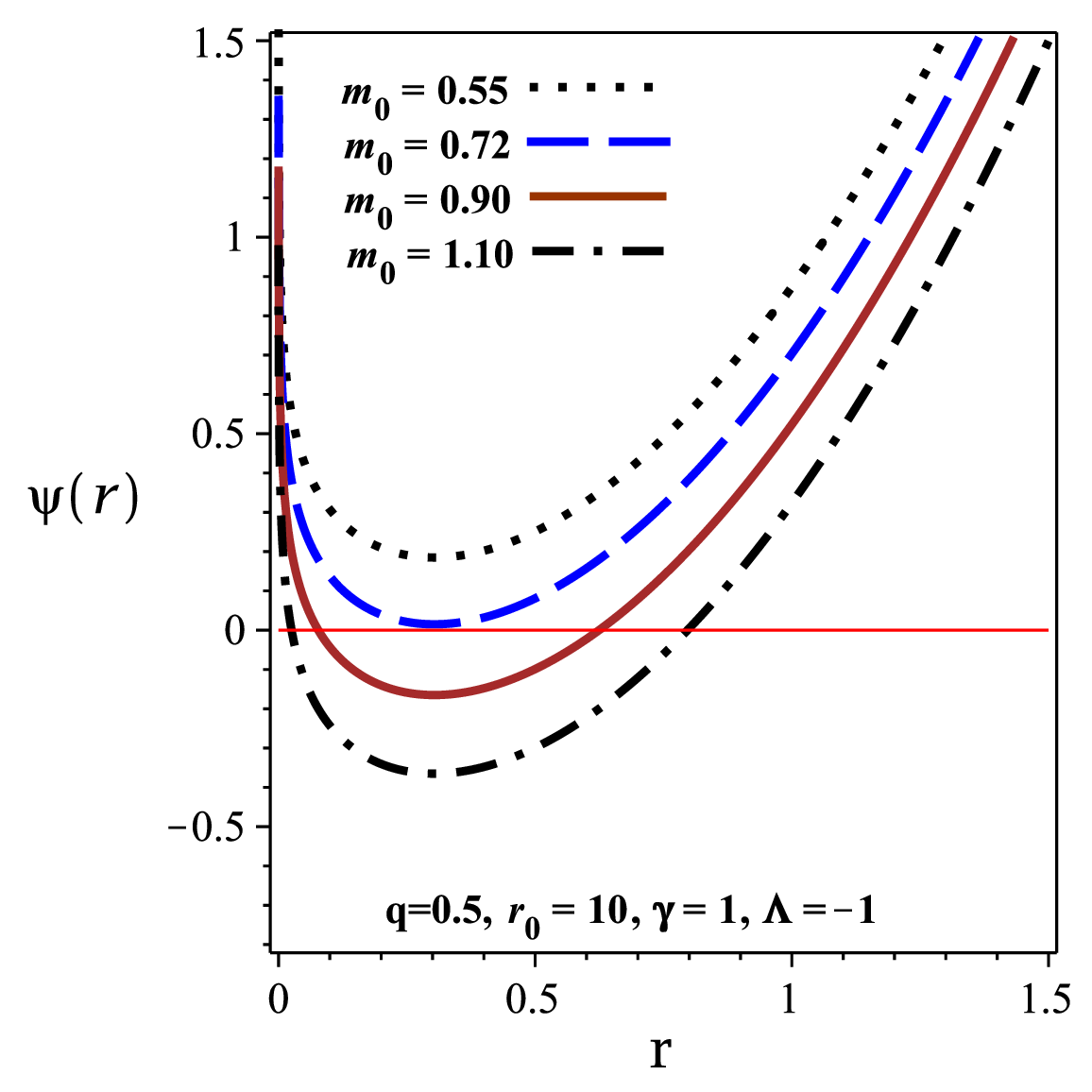} \includegraphics[width=0.4%
\linewidth]{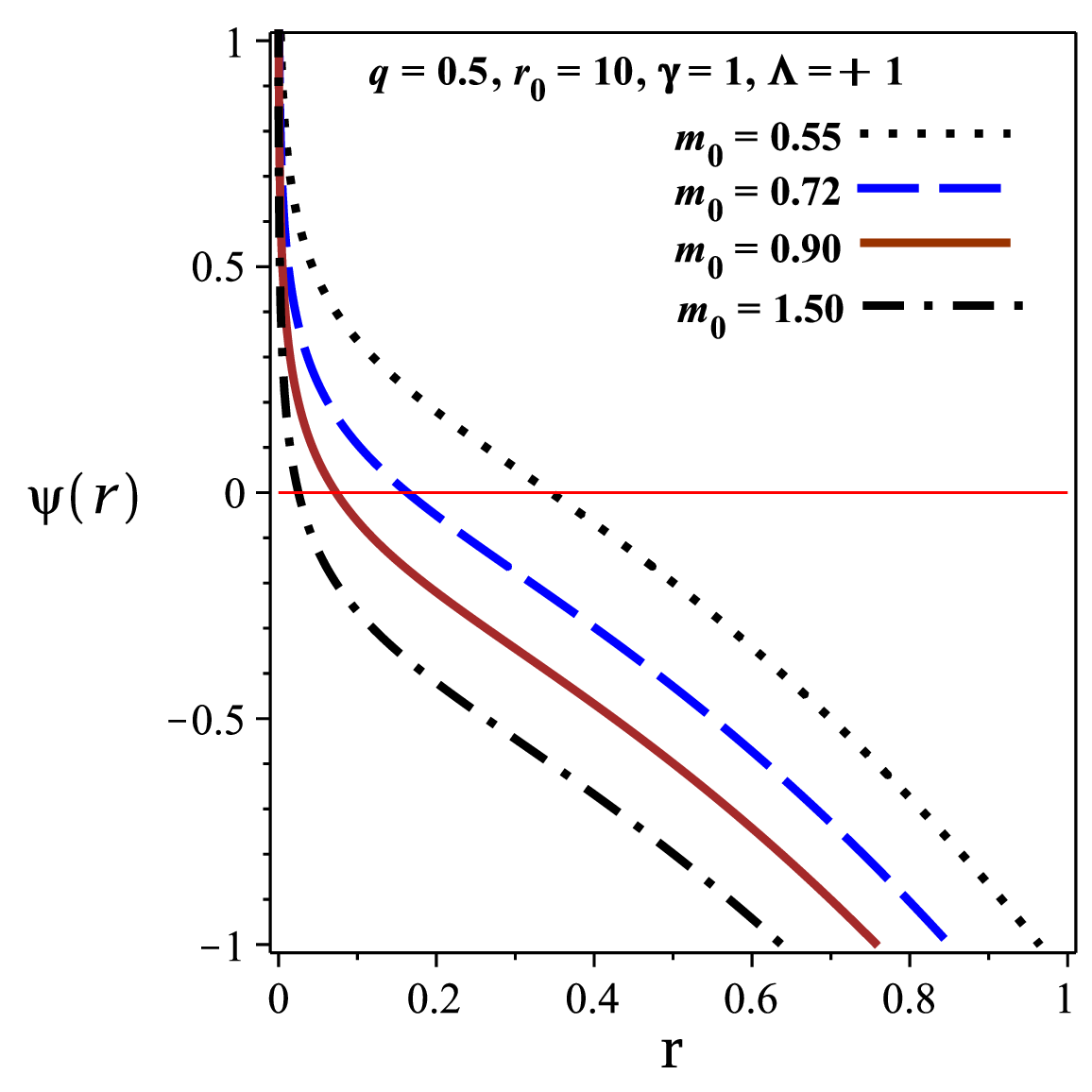} \newline
\includegraphics[width=0.4\linewidth]{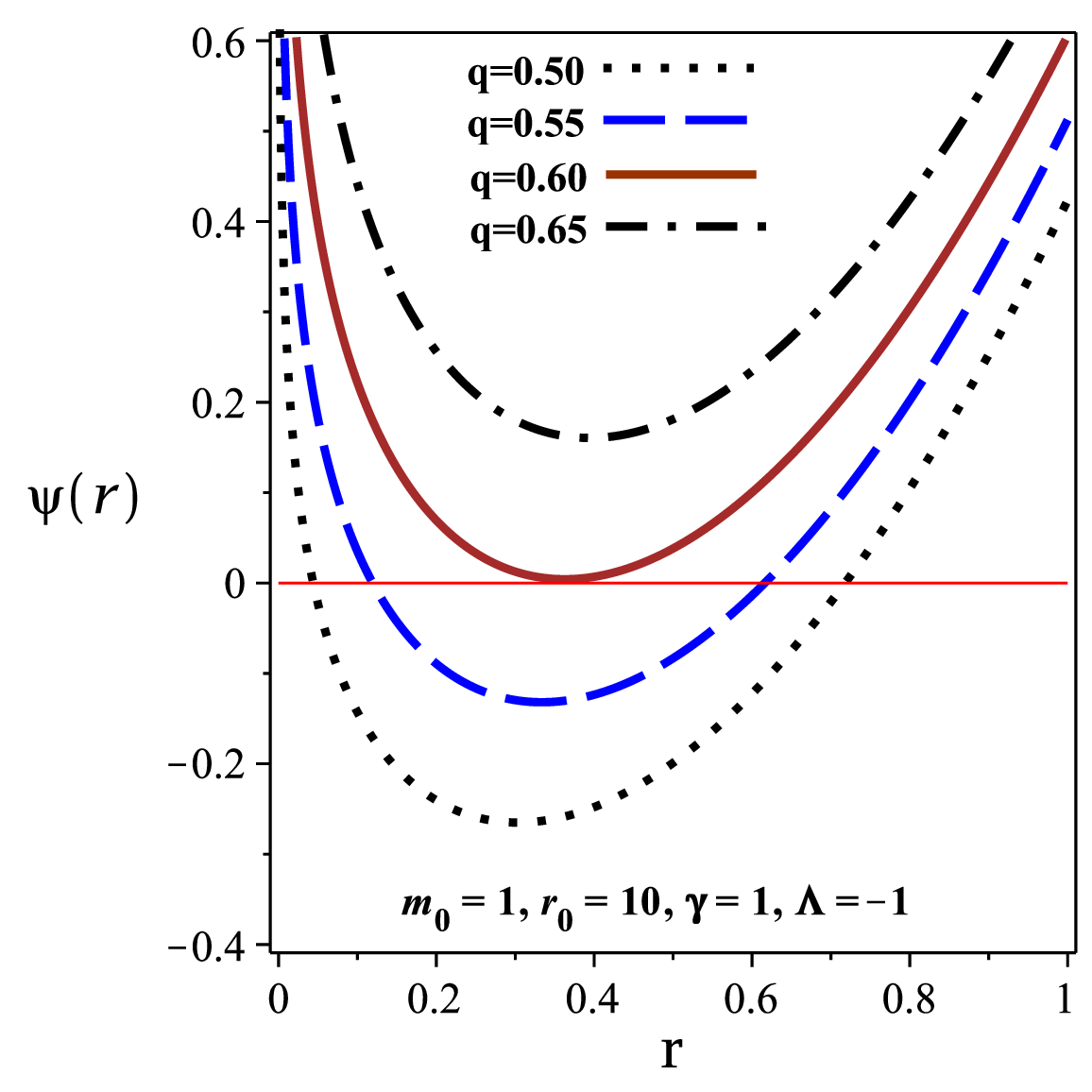} \includegraphics[width=0.4%
\linewidth]{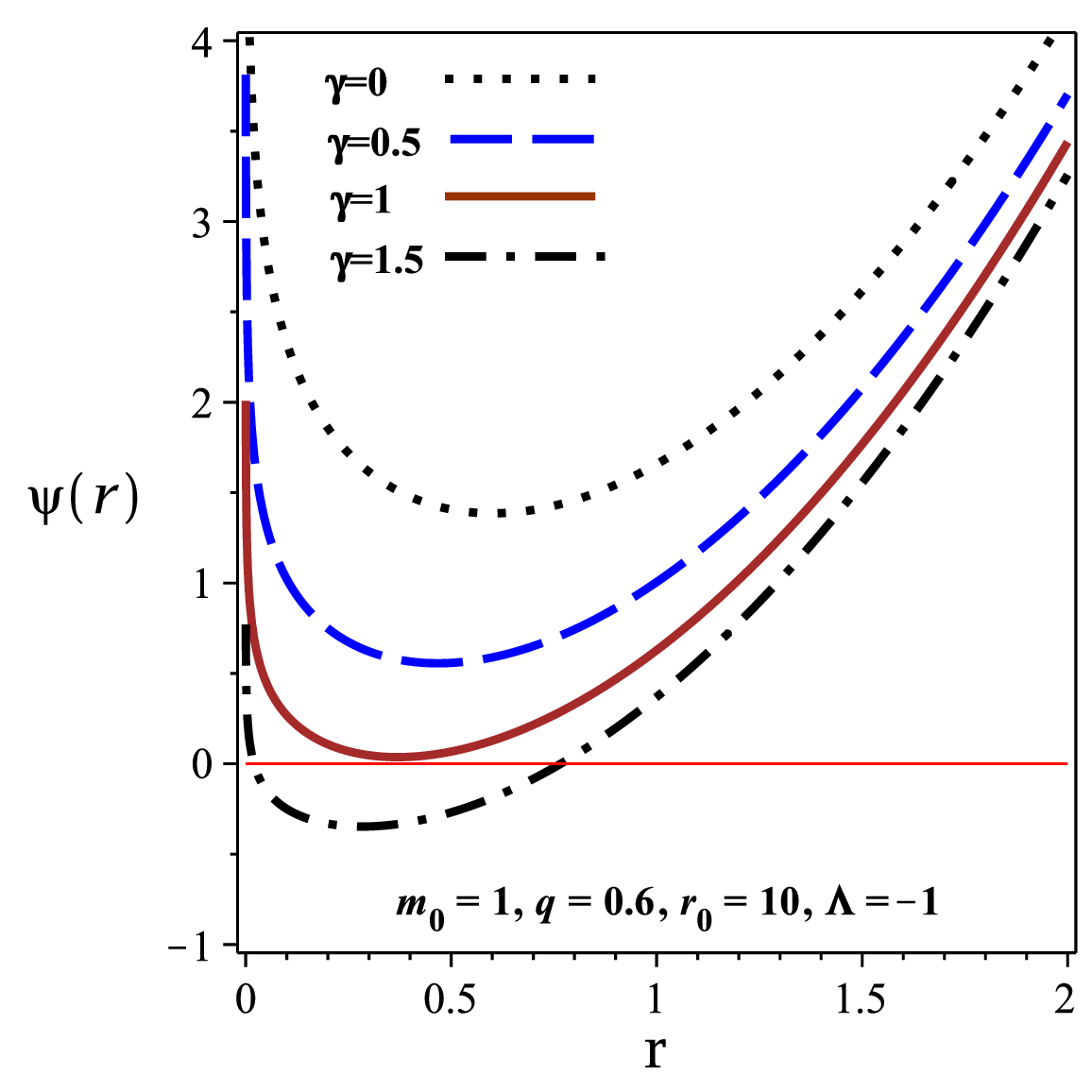} \newline
\caption{$\protect\psi (r)$ versus $r$ (Eq. (\protect\ref{f(r)EMMax})) for
different values of parameters.}
\label{Fig1}
\end{figure}

In order to get the real positive roots (or regular horizons) of the metric
function, we have plotted Fig. \ref{Fig1}. We can find two horizons (event
horizon and inner horizon), one extreme horizon, and also the naked
singularity (without horizon) for the obtained solutions. In other words,
the singularity is covered with at least one horizon (event horizon). So, we
can interpret these solutions as black holes. 

The effects of various parameters reveal that; i) by fixing the parameters
of $q$, $r_{0}$, $\gamma $, and $\Lambda $, the massive black holes have two
roots in which one of them is related to the event horizon (see the up-left
panel in Fig. \ref{Fig1}). ii) dS BTZ black hole does not include the event
horizon (see the up-right panel in Fig. \ref{Fig1}). So, the obtained
solution (\ref{f(r)EMMax}) cannot be a black hole when $\Lambda >0$. iii)
small electrical charge BTZ black holes include two roots, inner horizon,
and outer (event) horizon (see the down-left panel in Fig. \ref{Fig1}). iv)
the effect of ModMax parameter indicates that by increasing $\gamma $, the
number of root increases when other parameters are fixed (see the down-right
panel in Fig. \ref{Fig1}).

As a result, small charged AdS BTZ black holes with large values of the
ModMax parameter ($\gamma $) and mass ($m_{0}$) have two roots, which are
inner root and event horizon, respectively.

\section{Thermodynamic Quantities}

We obtain the thermodynamic quantities of BTZ-ModMax black
holes in this section.

The Hawking temperature is defined as 
\begin{equation}
T=\frac{\kappa }{2\pi }=\frac{\sqrt{\frac{-1}{2}\left( \nabla _{\mu }\chi
_{\nu }\right) \left( \nabla ^{\mu }\chi ^{\nu }\right) }}{2\pi },
\end{equation}
where $\chi =\partial _{t}$\ is the Killing vector, and also $\kappa $ is
the surface gravity. Using the metric (\ref{metric}), we can get the surface
gravity $\kappa $, which is given by $\kappa =\left. \frac{\psi ^{\prime }(r)%
}{2}\right\vert _{r=r_{+}}$, in which $r_{+}$ is the outer (event) horizon.
So the Hawking temperature of BTZ-ModMax black hole is 
\begin{equation}
T=\frac{\left. \psi ^{\prime }(r)\right\vert _{r=r_{+}}}{4\pi }=-\frac{%
\Lambda r_{+}}{2\pi }-\frac{q^{2}}{2\pi r_{+}}e^{-\gamma }.  \label{TotalTT}
\end{equation}

As one can see, the temperature depends on the cosmological constant ($%
\Lambda $), the electrical charge ($q$), and the parameter of ModMax ($%
\gamma $). The cosmological term has a positive effect on temperature due to
the existence of black hole solutions for the AdS case (i.e., $\Lambda <0$).
Therefore, the Hawking temperature is an increasing function of the
cosmological constant. In addition, by increasing $q$, the temperature
decreases because the electrical charge has a negative effect on $T$. By
increasing $\gamma $, the effect of the electrical charge disappears, and
the temperature will be independent of $q$. On the other hand, the roots of
temperature determine the bound points because, from a classical
thermodynamics point of view, the negative of temperature are devoted to
non-physical solutions. So, the roots of temperature can separate physical
from non-physical solutions. We find the roots of the temperature, which are 
\begin{equation}
r_{\pm }|_{T=0}=\frac{\pm q}{\sqrt{-\Lambda e^{\gamma }}},
\end{equation}%
which indicates that there is only one positive root ($r_{+}$). Also, a real
root for the temperature reveals that the cosmological constant has to be
negative $\Lambda <0$. This root (or the bound point) depends on three
parameters of our system, i.e., $\Lambda $, $q$, and $\gamma $. This bound
point decreases (increases) by increasing $\left\vert \Lambda \right\vert $
and $\gamma $ ($q$).

\begin{figure}[tbph]
\centering
\includegraphics[width=0.4\linewidth]{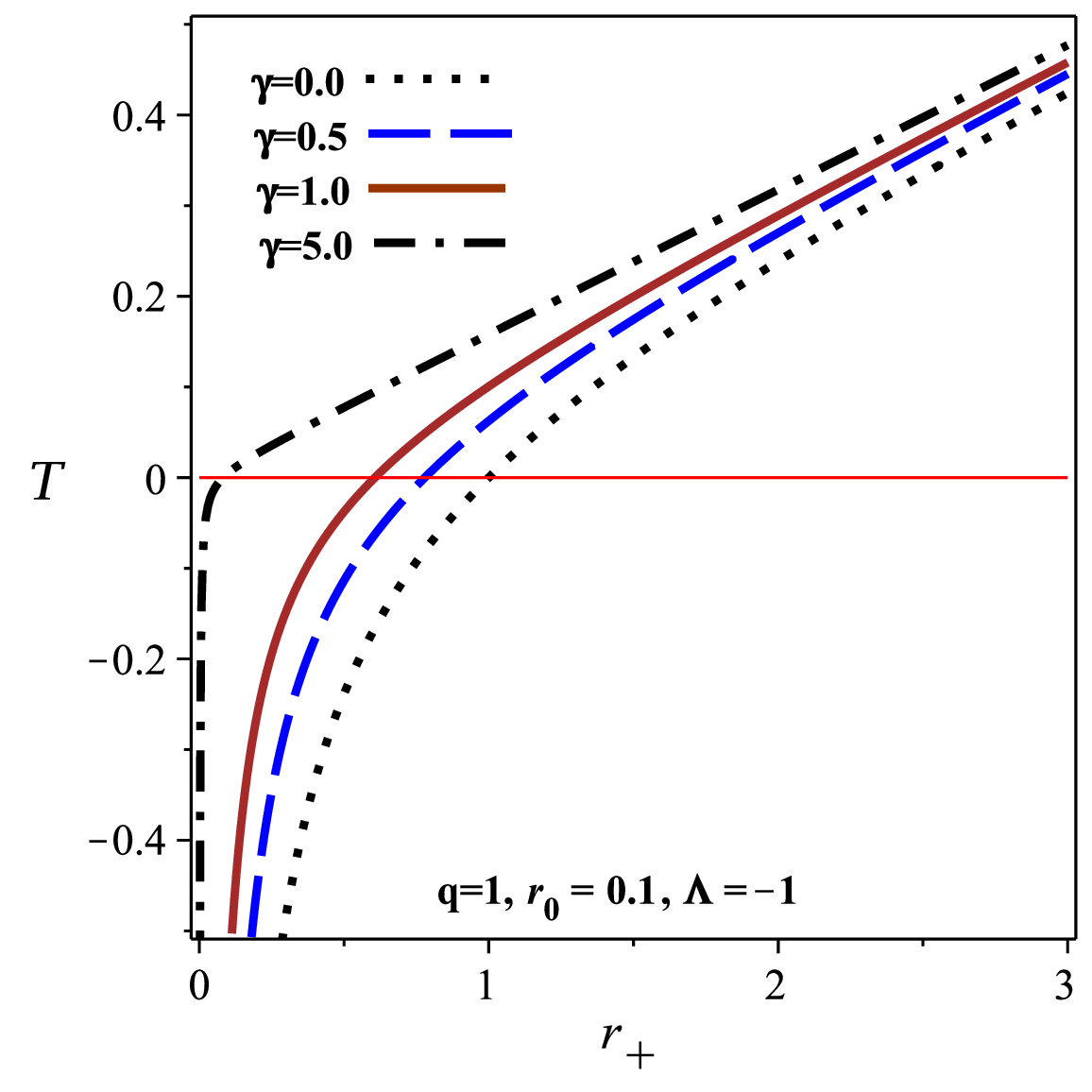}
\caption{$T$ versus $r_{+}$ for different values of parameters.}
\label{Fig2}
\end{figure}

In the high energy limit of the temperature is given by 
\begin{equation}
\lim_{very\text{ }small\text{ }r_{+}}T\propto -\frac{q^{2}}{2\pi r_{+}}%
e^{-\gamma },
\end{equation}%
which indicates that the ModMax parameter plays a significant role in the
high energy limit of the temperature. In other words, by increasing $\gamma $%
, this limit decreases, and finally, it will be zero (see Fig. \ref{Fig2},
for more details).

For the asymptotic limit of the temperature, we have 
\begin{equation}
\lim_{very\text{ \textit{large} }r_{+}}T\propto -\frac{\Lambda r_{+}}{2\pi },
\end{equation}%
which is dependent on the cosmological constant. According to this fact, the
cosmological constant is negative, so the asymptotic limit of the
temperature will be positive. Indeed, the temperature of very large
BTZ-ModMax black holes is always positive when $\Lambda <0$.

The electric charge, $Q$ is obtained by using Gauss's law in the following
form 
\begin{equation}
Q=\frac{q}{2}e^{-\gamma },  \label{TotalQ}
\end{equation}%
where depends on the ModMax parameter.

In Einstein's theory of gravity, the entropy of black holes ($S$) can be
extracted using the area law. In other words, we can obtain the entropy as a
quarter of the event horizon area $S=\frac{A}{4}$. Considering the metric (%
\ref{metric}), we can get the event horizon area ($A$) which is given by 
\begin{equation}
A=\int_{0}^{2\pi }\sqrt{g_{\varphi \varphi }}d\varphi =\left( \int_{0}^{2\pi
}d\varphi \right) r_{+}=2\pi r_{+},
\end{equation}%
so the entropy of BTZ-ModMax black holes is 
\begin{equation}
S=\frac{\pi r_{+}}{2}.  \label{TotalS}
\end{equation}

Applying the Hamiltonian approach and also the counterterm method, we are
able to obtain the total mass of solutions, which leads to 
\begin{equation}
M=\frac{m_{0}}{8},  \label{TotalM}
\end{equation}%
where $m_{0}$ is geometrical mass and gets from the metric function (\ref%
{f(r)EMMax}) on the horizon ($\psi \left( r=r_{+}\right) =0$), which leads
to 
\begin{equation}
m_{0}=-\Lambda r_{+}^{2}-2q^{2}e^{-\gamma }\ln \left( \frac{r_{+}}{r_{0}}%
\right) ,  \notag
\end{equation}%
so, the total mass turns to 
\begin{equation}
M=-\frac{\Lambda r_{+}^{2}}{8}-\frac{q^{2}e^{-\gamma }}{4}\ln \left( \frac{%
r_{+}}{r_{0}}\right) .  \label{MM1}
\end{equation}

Similar to the temperature (\ref{TotalTT}), the total mass depends on the
cosmological constant, the electrical charge, and the ModMax parameter. On
the other hand, it was discussed that the total mass of black holes might be
considered as the internal energy of a system. It is clear that the internal
energy (on total mass) is always positive for large black holes. To study
the effects of the MadMax parameter on the internal energy of our system, we
plot the total mass (\ref{MM1}) versus the radius of black holes in Fig. \ref%
{Fig3}.

\begin{figure}[tbph]
\centering
\includegraphics[width=0.4\linewidth]{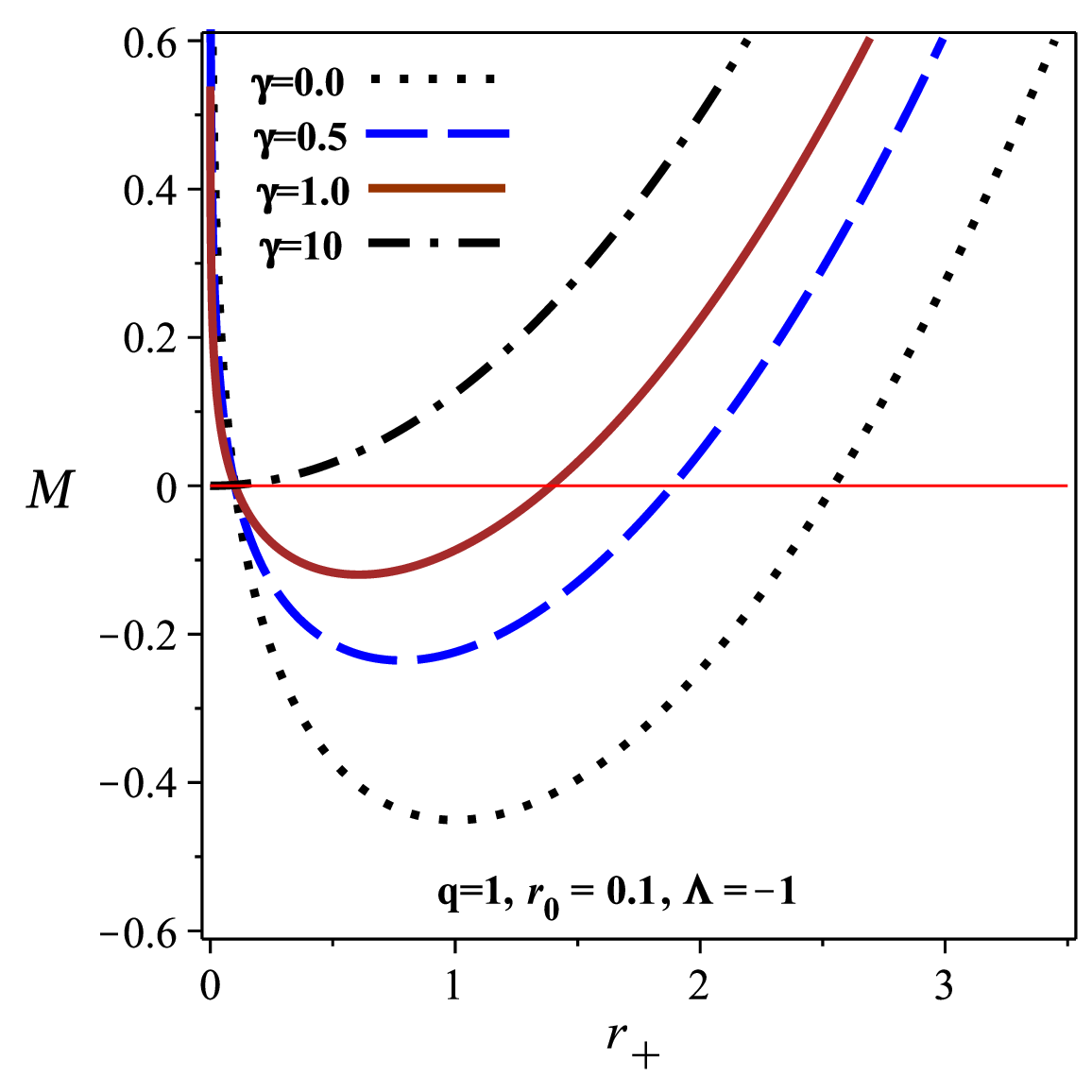}
\caption{$M$ versus $r_{+}$ for different values of parameters.}
\label{Fig3}
\end{figure}

Our findings indicate that by increasing $\gamma $, the negative values of
the mass disappear, and we encounter the positive mass everywhere. Indeed,
by increasing the ModMax parameter, the negative area decreases, and in the
limit of very large values of $\gamma $, there is no negative area for the
total mass.

We can get the electric potential, $U$, by the difference of gauge potential
between the reference and the horizon, which yields \cite%
{Cvetic1999,Caldarelli2000} 
\begin{equation}
U=A_{\mu }\chi ^{\mu }\left\vert _{r\rightarrow reference}\right. -A_{\mu
}\chi ^{\mu }\left\vert _{r\rightarrow r_{+}}\right. =-q\ln \left( \frac{%
r_{+}}{r_{0}}\right) .  \label{TotalU}
\end{equation}

Now, we are able to check the relation of the first law of thermodynamics.
Using the obtained thermodynamic quantities such as electric charge (\ref%
{TotalQ}), entropy (\ref{TotalS}), and mass (\ref{TotalM}), and after some
calculations, the first law of black hole thermodynamics is in the following
relation 
\begin{equation}
dM=TdS+UdQ,
\end{equation}%
where $S$ and $Q$, respectively, are the temperature and the electric
potential in the following forms 
\begin{eqnarray}
T &=&\left( \frac{\partial M}{\partial S}\right) _{Q},  \label{TU1} \\
&&  \notag \\
U &=&\left( \frac{\partial M}{\partial Q}\right) _{S}.  \label{TU2}
\end{eqnarray}

\section{Thermal Stability}

Here, we would like to study the thermal stability of BTZ-ModMax black holes
in the context of the canonical ensemble. In this regard, the heat capacity
and the Helmholtz free energy play a significant role in determining thermal
stability. Using the heat capacity and the Helmholtz free energy, we can
evaluate the local and global stability, respectively. So, we discuss the
thermal stability of BTZ black holes by using the heat capacity and the
Helmholtz free energy.

\subsection{\textbf{Local stability}}

By studying the heat capacity, we can extract two significant properties of
the solutions, which are phase transition points, and the thermal stability
of the solutions. Indeed, the phase transition points are related to the
divergences of the heat capacity because these divergences belong to
where the under-studying system goes under phase transitions. Also, the signature of heat capacity (i.e. bound point) determines the
thermal instability/stability of the system in the canonical
ensemble. In other words, the positivity (the negative) of heat capacity
reveals that the black hole is in a thermally (un)stable state. So, we have to find bound (roots of the heat capacity is where the sign
of temperature is changed) and phase transition (divergences point of the
heat capacity) points.

The heat capacity with fixed charge is defined by 
\begin{equation}
C_{Q}=\frac{T}{\left( \frac{\partial T}{\partial S}\right) _{Q}}=\frac{%
\left( \frac{\partial M\left( S,Q\right) }{\partial S}\right) _{Q}}{\left( 
\frac{\partial ^{2}M\left( S,Q\right) }{\partial S^{2}}\right) _{Q}}.
\label{C}
\end{equation}%
to extract the heat capacity, we re-write the Hawking temperature (\ref%
{TotalTT}) and the total mass (\ref{MM1}) of BTZ-ModMax black hole in terms
of the electrical charge (\ref{TotalQ}), and the entropy (\ref{TotalS}), in
the following forms 
\begin{eqnarray}
T &=&\left( \frac{\partial M\left( S,Q\right) }{\partial S}\right) _{Q}=%
\frac{-\Lambda S}{\pi ^{2}}-\frac{Q^{2}e^{\gamma }}{S},  \label{TM} \\
&&  \notag \\
M\left( S,Q\right) &=&\frac{\Lambda S^{2}}{2\pi ^{2}}+Q^{2}e^{\gamma }\ln
\left( \frac{2S}{\pi r_{0}}\right) .  \label{MSQ}
\end{eqnarray}

Now, we can get the heat capacity by using Eqs. (\ref{C})-(\ref{MSQ}), which
is 
\begin{equation}
C_{Q}=\frac{\left( \Lambda S^{2}+\pi ^{2}Q^{2}e^{\gamma }\right) S}{\Lambda
S^{2}-\pi ^{2}Q^{2}e^{\gamma }},  \label{heatC}
\end{equation}%
which indicates the heat capacity depends on the cosmological constant,
electrical charge, and the ModMax parameter. To find the bound and phase
transition points, we have to solve the following relations 
\begin{equation}
\left\{ 
\begin{array}{ccc}
\left( \frac{\partial M\left( S,Q\right) }{\partial S}\right) _{Q}=0, &  & 
\text{bound points} \\ 
&  &  \\ 
\left( \frac{\partial ^{2}M\left( S,Q\right) }{\partial S^{2}}\right) _{Q}=0,
&  & \text{phase transition points}%
\end{array}%
\right. .  \label{PhysBound}
\end{equation}

\begin{figure}[tbph]
\centering
\includegraphics[width=0.4\linewidth]{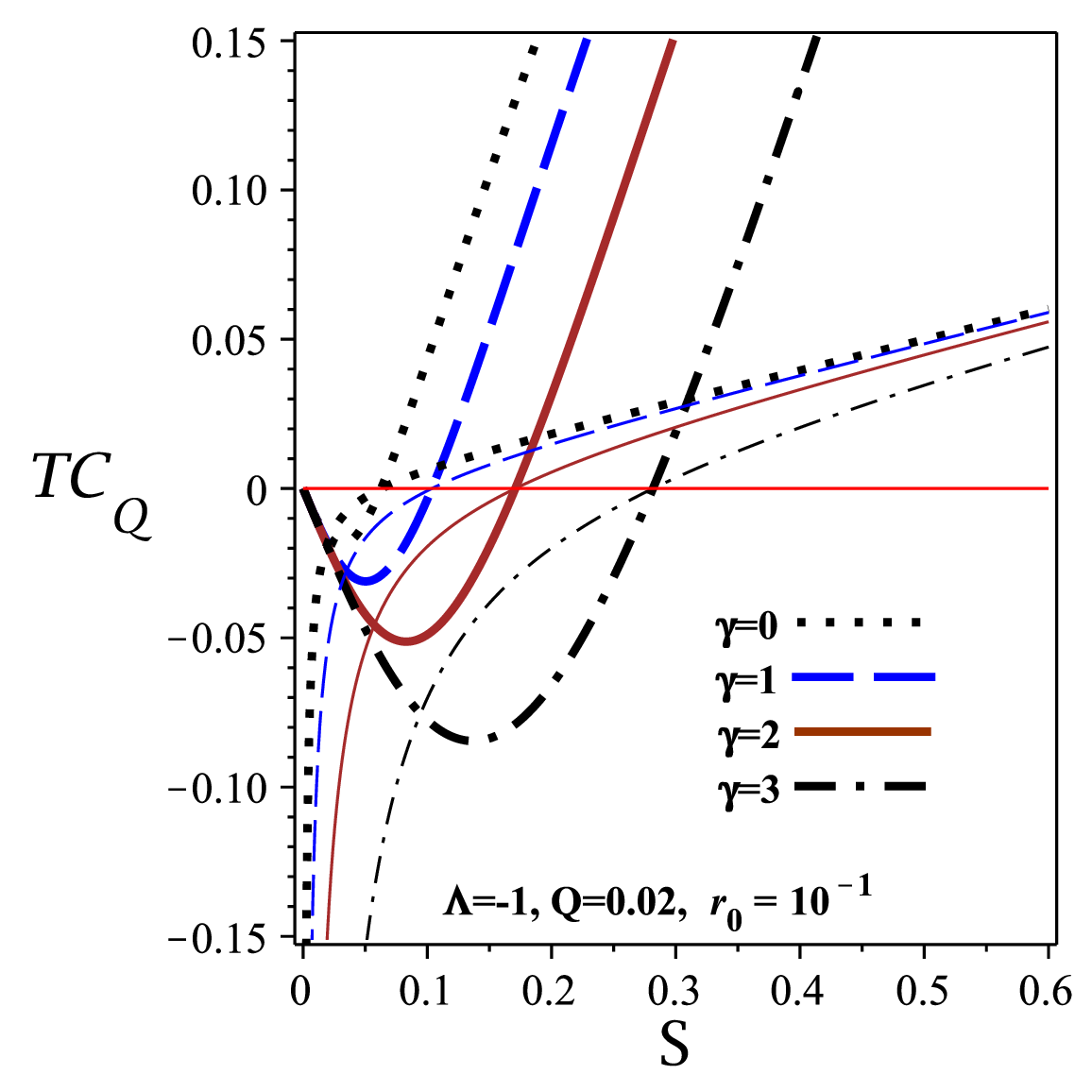}
\caption{Heat capacity $C_{Q}$ (thick lines), and the Hawking temperature $T$
(thin lines) versus $S$ for different values of the ModMax parameter.}
\label{Fig4}
\end{figure}

We first get the bound point by solving $T=0$. For this purpose, we consider
Eq. (\ref{TM}) and solve it in terms of the entropy, which leads to 
\begin{equation}
S_{root}=\pi Q\sqrt{\frac{e^{\gamma }}{-\Lambda }},  \label{RootT}
\end{equation}%
which states that there is one real root (or one bound point) for the heat
capacity. In other words, there are two different (thermal stable/unstable)
areas for the system that are before and after this bound point. It is
notable that the bound point depends on the electrical charge, the
cosmological constant, and the ModMax parameter. On the other hand, the root
of temperature reveals a limitation point. Indeed, the root of the
temperature separates physical (i.e. positive temperature) from non-physical
solutions (i.e. negative temperature). For determining the thermal
stability/instability and physical black holes, we plot the heat capacity
versus entropy in Fig. \ref{Fig4}.

Our results in Fig. \ref{Fig4} show that there are two different behaviors
for the heat capacity which are; i) in the range $S<S_{root}$, black holes
are not physical objects because the temperature is negative. Since the heat
capacity is negative in this area, our system is thermally unstable. ii) in
the range $S>S_{root}$, the temperature and the heat capacity are positive.
So, black holes in this area are physical and stable objects. The effect of
the ModMax parameter reveals that the physical and thermal stable area
decrease by increasing $\gamma $ (see Fig. \ref{Fig4}, for more details). As
a result, AdS BTZ-ModMax black holes with a large radius (or entropy) are
physical and thermal stable objects.

\subsection{\textbf{Global stability}}

The Helmholtz free energy in the usual case of thermodynamics, is defined as 
\begin{equation}
F=U-TS,
\end{equation}%
where in the context of the black holes, we can consider $U=M$. So, in a
canonical ensemble with a fixed charge $Q$, the Helmholtz free energy is
given by 
\begin{equation}
F(T,Q)=M\left( S,Q\right) -TS,  \label{free}
\end{equation}%
after some calculation, we get 
\begin{equation}
F=-Q^{2}e^{\gamma }\left( \ln \left( \frac{2S}{\pi r_{0}}\right) -1\right) +%
\frac{\Lambda S^{2}}{2\pi ^{2}}.
\end{equation}

\begin{figure}[tbph]
\centering
\includegraphics[width=0.4\linewidth]{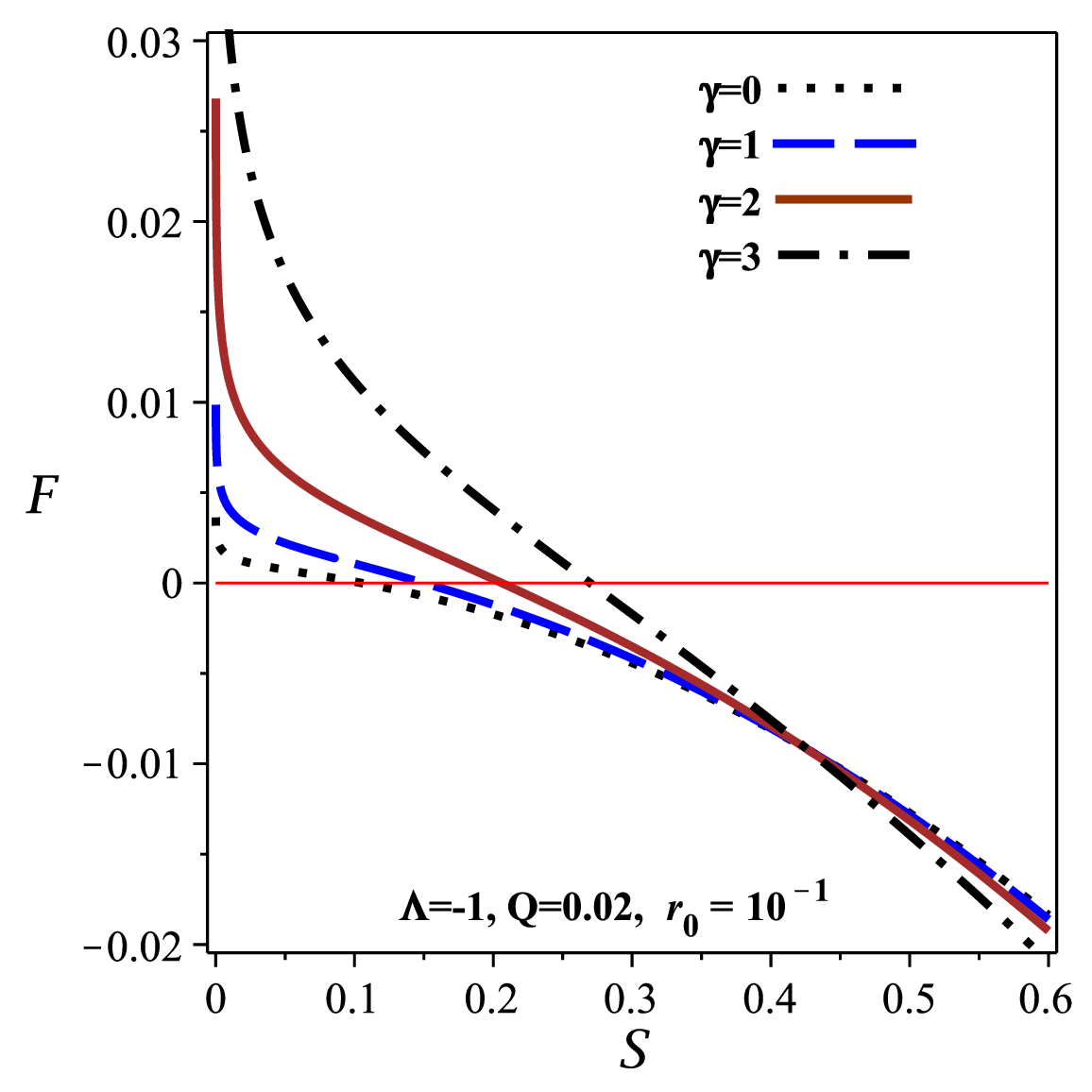}
\caption{Helmholtz free energy $F$ versus $S$ for different values of the
ModMax parameter.}
\label{Fig5}
\end{figure}

It is notable that in the context of the canonical ensemble, $F<0$ (i.e. the
negative of the Helmholtz free energy) determines the global stability of a
thermodynamic system. In this regard, we find one real root of the Helmholtz
free energy in the following form 
\begin{equation}
S_{root_{F}}=\frac{\pi r_{0}}{2}e^{-\frac{LambertW\left( \frac{-\Lambda
r_{0}^{2}}{4Q^{2}e^{\gamma -2}}\right) }{2}+1},
\end{equation}%
which indicates that this root depends on all of our system's parameters. To
determine the global stability, we plot the Helmholtz free energy versus
entropy in Fig. \ref{Fig5}.

Our results in Fig. \ref{Fig5}, indicate that there are two different areas,
which are; i) AdS BTZ-ModMax black holes do not satisfy the global stability
in the range $S<S_{root_{F}}$ because $F>0$ in this area. ii) In the range $%
S>S_{root_{F}}$, the Helmholtz free energy is negative, so the black holes
satisfy the global stability.

The effect of the ModMax parameter indicates that the global stable area
decreases by increasing $\gamma $ (similar to the heat capacity). As a
result, AdS BTZ-ModMax black holes with large radius (or entropy) are within
the global stable area.

By comparing the local and global stabilities of AdS BTZ-ModMax black holes,
simultaneously, we find that the AdS BTZ-ModMax black holes with large radii
(or entropy) satisfy the local and global stability conditions,
simultaneously.

\section{Conclusions}

In this paper, we obtained exact analytical three-dimensional solutions in
the presence of a new NED model, which is known as ModMax NED. Our analysis
indicated that this solution belongs to the black hole solution which
included a singularity at $r=0$, which is covered by an event horizon for
the AdS case. Also, we showed that although the ModMax parameter could
remove the singularity of the electrical field at $r=0$, however, it did not
remove the curvature singularity at $r=0$. Next, we evaluated the effects of
various parameters on the root of the metric function. Our analysis in Fig. %
\ref{Fig1} revealed that small charged AdS BTZ black holes with big values
of the ModMax parameter ($\gamma $) and mass ($m_{0}$) had two roots, which
were related to inner root and event horizon, respectively. Moreover, by
increasing (decreasing) $\gamma $ and $m_{0}$ ($q$), the radius of the event
horizon increases, and we encounter large black holes.

We calculated the conserved and thermodynamic quantities of the solution
such as Hawking temperature, electric charge, electric potential, entropy, and mass. Our analysis of the Hawking temperature indicated that it depended on the cosmological constant ($\Lambda $), the electrical charge ($%
q $), and the parameter of ModMax ($\gamma $). $\Lambda $ ($q$) had a
positive (negative) effect on temperature. Indeed, the temperature was an
increasing (decreasing) function of the cosmological constant (the
electrical charge). In addition, by increasing $\gamma $, the Hawking temperature was independent of $q$. We studied the high energy and asymptotic limit of the Hawking temperature. Our analysis indicated that the high energy limit of the temperature was dependent on $\gamma $, whereas the asymptotic limit of the temperature was dependent on the cosmological constant. Another interesting thermodynamics quantity was related to the total mass of black holes because this quantity gives us information about internal energy. Our findings in Fig. \ref{Fig3} indicated that by increasing the ModMax parameter, the negative area of the total mass decreased, and in the limit of very large values of $\gamma $, there was no negative area for the total mass. Then, we checked the validity of the first law of thermodynamics for BTZ-ModMax black holes.

To study the thermal stability (i.e. the local and global stabilities) of BTZ-ModMax black holes in the canonical ensemble, we evaluated simultaneously the heat capacity and the Helmholtz free energy for these black holes.

Our analysis of the heat capacity indicated that there was one bound point ($%
S_{root}$) which was dependent on the electrical charge, the cosmological
constant, and the ModMax parameter. This bound point separated two different
behaviors for the heat capacity, which were before and after this point. In
other words, before the bound point, i.e., in the range $S<S_{root}$, black
holes were not physical and stable objects because the temperature and the
heat capacity were negative. After the bound point, i.e., in the range $%
S>S_{root}$, the temperature and the heat capacity were positive. Therefore,
large black holes were physical and stable objects. The effect of the ModMax
parameter on the bound point in Fig. \ref{Fig4} revealed that the physical
and thermal stable area decreased by increasing $\gamma $.

We studied the Helmholtz free energy in order to evaluate the global
stability. Our results in Fig. \ref{Fig5} indicated that there was the same
behavior for the Helmholtz free energy, similar to the heat capacity. There
was one real root for the Helmholtz free energy in which before and after
this root, the Helmholtz free energy was positive and negative,
respectively. Indeed, the large black holes satisfied the global stability
condition. In addition, the effect of the ModMax parameter showed that the
global stable area decreased by increasing $\gamma $.

Finally, we compared the local and global stability, together. We found that
the AdS BTZ-ModMax black hole with a large radius (or entropy) could satisfy
the local and global stability conditions, simultaneously.

\begin{acknowledgements}
	
I would like to thank University of Mazandaran.
\end{acknowledgements}

\end{document}